\documentclass[aps,prb,superscriptaddress,showpacs,twocolumn]{revtex4}
\usepackage{graphicx}
\usepackage{amssymb}
\usepackage{amsmath}

\newcommand{\bk}{{\bf k}}
\newcommand{\bq}{{\bf q}}

\newcommand{\br}{{\bf r}}

\renewcommand{\Im}{{\mathop{\rm{Im}}\nolimits\,}}
\renewcommand{\Re}{{\mathop{\rm{Re}}\nolimits\,}}

\newcommand{\sech}{{\mathop{\rm{sech}}\nolimits\,}}

\begin{document}

\bibliographystyle{apsrev}

\title{Effects of proximity to an electronic topological transition\\
   on normal state transport properties of the high-$T_c$ superconductors}

\author{G. G. N. Angilella}
\author{R. Pucci}
\affiliation{Dipartimento di Fisica e Astronomia, Universit\`a di
   Catania,\\ and Istituto Nazionale per la Fisica della Materia,
   UdR di Catania,\\ Via S. Sofia, 64, I-95123 Catania, Italy}
\author{A. A. Varlamov}
\affiliation{Istituto Nazionale per la Fisica della Materia,
   UdR ``Tor Vergata''/Coherentia,\\
   Via del Politecnico, 1, I-00133 Roma, Italy}
\author{F. Onufrieva}
\affiliation{Laboratoire L\'eon Brillouin, CE-Saclay, F-91191
   Gif-sur-Yvette, France}

\date{January 13, 2003}

\begin{abstract}
\medskip
Within the time dependent Ginzburg-Landau theory, the effects of the
   superconducting fluctuations on the transport properties above the
   critical temperature are characterized by a non-zero imaginary part
   of the relaxation rate $\gamma$ of the order parameter.
Here, we evaluate $\Im\gamma$ for an anisotropic dispersion relation
   typical of the high-$T_c$ cuprate superconductors (HTS),
   characterized by a proximity to an electronic topological
   transition (ETT).
We find that $\Im\gamma$ abruptly changes sign at the ETT
   as a function of doping, in agreement with the universal behavior
   of the HTS.
We also find that an increase of the in-plane anisotropy, as is given
   by a non-zero value of the next-nearest to nearest hopping ratio
   $r=t^\prime /t$, increases the value of $|\Im\gamma|$ close to the
   ETT, as well as its singular behavior at low temperature,
   therefore enhancing the effect of superconducting fluctuations.
Such a result is in qualitative agreement with the available data for
   the excess Hall conductivity for several cuprates and cuprate
   superlattices.
\\
\end{abstract} 

\maketitle

\section{Introduction}

The appearance of superconducting fluctuations above the critical
   temperature $T_c$ leads to precursor effects of the superconducting
   phase occurring already above $T_c$.
Due to their short coherence length, the discovery of
   the high-$T_c$ cuprate superconductors (HTS) made the fluctuation
   regime experimentally accessible over a relatively wide temperature
   range above $T_c$ \cite{Larkin:02}. 
Superconducting fluctuations manifest themselves in the singular temperature
   dependence of thermodynamic properties, such as the specific heat
   and the susceptibility, and of several transport properties
   (see Refs.~\onlinecite{Varlamov:99,Larkin:02} for recent reviews).
In particular, the influence of superconducting fluctuations on the
   Ettinghausen effect \cite{Ullah:90}, the Nernst effect, the
   thermopower, the electrical conductivity, and the Hall conductivity
   \cite{Ullah:91} have been considered within the time-dependent
   Ginzburg-Landau (TDGL) theory for a layered superconductor in a
   magnetic field near $T_c$.
A numerical approach within the fluctuation exchange (FLEX)
   approximation to the theory of 
   electric transport in the normal state of the high-$T_c$ cuprates has been
   developed by Yanase \emph{et al.}
   \cite{Yanase:99,Yanase:00,Yanase:01,Yanase:02}. 

The effect of fluctuations on the transport properties of the high-$T_c$
   superconductors can contribute to a better understanding of the
   unconventional properties of their normal state.
Recent experimental studies of the Nernst effect in underdoped cuprates
   have demonstrated a sizeable Nernst coefficient in the normal state
   both at high temperature and in high magnetic fields
   \cite{Xu:00,Capan:02,Wang:02a}.
Such findings have been interpreted as an effect of precursor pairing
   above $T_c$ in the pseudogap region, as well as of quantum
   superconducting fluctuations \cite{Ikeda:02}.

In the case of the Hall effect, superconducting fluctuations induce a
   characteristic deviation from the normal state temperature
   dependence of the Hall conductivity above $T_c$ (Hall anomaly) \cite{Iye:92}.
In particular, the fluctuation Hall conductivity $\Delta\sigma_{xy}$ has been
   evaluated within the TDGL theory \cite{Dorsey:92,Ullah:91}, and it
   has been shown that a Hall sign reversal takes place below $T_c$.
The value and sign of $\Delta\sigma_{xy}$ strongly depends on the
   electronic structure of the material under consideration and, in
   particular, on the topology of its Fermi surface.
It is well known that $\Delta\sigma_{xy}$ arises as a result of an
   electron-hole asymmetry in the band structure \cite{Fukuyama:71}.
Recently, on the basis of the general requirement of gauge invariance of
   the TDGL equations, it has been shown that the sign of
   $\Delta\sigma_{xy}$ is determined by $\partial\ln T_c /
   \partial\ln\mu$, where $\mu$ is the chemical potential
   \cite{Aronov:95}.
More recently, evidence for a universal behavior of the Hall
   conductivity as a function of doping has been reported in the
   cuprate superconductors \cite{Nagaoka:98}.

Given the relevance of the electronic structure in establishing the
   magnitude and sign of the fluctuation Hall effect, it is of obvious
   interest to study the effect of fluctuations on the transport
   properties of low-dimensional superconducting materials in the
   proximity of an electronic topological transition (ETT)
   \cite{Lifshitz:60,Varlamov:89,Blanter:94}.
An ETT consists of a change of topology of the Fermi surface, and may
   be induced by doping, as well as by changing the impurity
   concentration, or applying pressure or anisotropic stress.
In all such cases, one may introduce a critical parameter $z$,
   measuring the proximity to the ETT occurring at $z=0$.
In the case of quasi two-dimensional (2D) materials, such as the
   cuprates, the electronic band is locally characterized by a
   hyperbolic-like dispersion relation.
Therefore, one is particularly interested in the study of an ETT of
   the `neck disruption' kind, according to the original
   classification of I. M. Lifshitz \cite{Lifshitz:60}.

Some effects of an ETT (namely, the existence of a Van~Hove
   singularity in the density of states) on the superconducting
   properties of the cuprates are well known
   \cite{Pattnaik:92,Newns:92,Newns:92a,Markiewicz:97}.
Recently, it has been shown also that the effect of the proximity to
   an ETT is richer than having a Van~Hove singularity in the density
   of states, namely that the ETT is a specific quantum critical
   point.
This leads to the existence of several quantum critical regimes that
   can explain the observed anomalous properties of the high-$T_c$
   cuprates in the normal state
   \cite{Onufrieva:96,Onufrieva:99a,Onufrieva:99b,Onufrieva:00}.
Some of the present authors have recently investigated the dependence
   of such effects on some specific material properties, such as the
   next-nearest to nearest neighbors hopping ratio
   \cite{Angilella:01}, and anisotropic stress \cite{Angilella:02d}.
Concerning the normal state transport properties of a superconductor, the
   effect of the proximity to an ETT has been studied for the
   thermoelectric power in a quasi-2D metal \cite{Blanter:92}, and for
   the Nernst and the weak-field Hall effects for both 3D and quasi-2D
   metals \cite{Livanov:99}.

In this paper, we will study the anomalous Hall conductivity due to
   the superconducting fluctuations above $T_c$ for a quasi-2D
   superconductor close to an ETT.
The link between TDGL theory and the microscopic
   theory is provided by the relaxation rate $\gamma$ of the
   fluctuating superconducting order parameter.
In particular, a non-zero imaginary part of this quantity gives rise
   to a fluctuation contribution to the Hall effect.
Here, we will study $\Im\gamma$ as a function of the ETT parameter
   $z$ and temperature $T$, both numerically and analitically, for
   a realistic band dispersion typical of the high-$T_c$ cuprate
   compounds.
Close to the ETT, $\Im\gamma$ is characterized by a steep inflection
   point, surrounded by a minimum and a maximum, whose
   height increases with decreasing temperature.
In the presence of electron-hole symmetry, we will show that
   $\Im\gamma$ is an odd function of the ETT parameter $z$, and that
   $\Im\gamma$ vanishes and rapidly changes sign at the ETT point.
In the cuprate superconductors, electron-hole symmetry is usually
   destroyed by a non-zero next-nearest to nearest neighbors hopping
   ratio $r=t^\prime /t$ \cite{Pavarini:01}.
In this case, the peaks in $\Im\gamma$ around the ETT point have
   unequal heights, and we will show that their dependence on the hopping
   parameter $r$ is in qualitative agreement with the results of
   several fits against the fluctuation Hall conductivity data of
   various cuprates and cuprate superlattices.

The paper is organized as follows.
In Sec.~\ref{sec:Hall} we will briefly review the 
   TDGL theory of superconducting fluctuations and the microscopic
   results for the direct and indirect contributions to the excess
   Hall conductivity $\Delta\sigma_{xy}$.
In Sec.~\ref{sec:micro} we will outline the microscopic derivation of
   $\gamma$, and explicitly evaluate $\Im\gamma$ as a function of the
   chemical potential and temperature.
We will eventually summarize in Sec.~\ref{sec:conclusions}.

\section{Excess Hall conductivity}
\label{sec:Hall}

A phenomenological description of the fluctuation effects on the
   transport properties of a layered superconductor is based on the time
   dependent Ginzburg-Landau (TDGL) equation \cite{Larkin:02}:
\begin{equation}
-\gamma \left( \frac{\partial}{\partial t} + \frac{2ie}{\hbar c}
   \varphi \right) \psi_\ell (\br,t) = 
   \frac{\delta\mathcal{F}}{\delta\psi_\ell^\ast (\br,t)}
   + \zeta (\br,t).
\label{eq:TDGL}
\end{equation}
Here, $\psi_\ell (\br,t)$ is the fluctuating Ginzburg-Landau (GL)
   order parameter on 
   layer $\ell$, $\varphi$ the scalar
   potential of the electic field, and $\zeta(\br,t)$ is the Langevin
   force, taking into account for the order parameter dynamics.
In the case of a layered superconductor, the GL functional
   $\mathcal{F}$ within the Lawrence-Doniach model \cite{Lawrence:71}
   takes the form:
\begin{widetext}
\begin{equation}
\mathcal{F} = \sum_\ell \int d^2 \br \left[ a |\psi_\ell |^2 +
   \frac{1}{2} b |\psi_\ell |^4 
+ \frac{\hbar^2}{4m} \left| \left(\nabla_\parallel -\frac{2ie}{\hbar c}
   {\bf A}_\parallel \right) \psi_\ell \right|^2 + \mathcal{J} |\psi_{\ell+1} -
   \psi_\ell |^2 \right],
\end{equation}
\end{widetext}
where $a$ and $b$ are the usual GL coefficients, ${\bf A}_\parallel$
   is the vector 
   potential of a magnetic field perpendicular to the layers, and
   $\mathcal{J}$ characterizes 
   the Josephson coupling between adjacent planes
   \cite{Larkin:02}.

In Eq.~(\ref{eq:TDGL}), the complex quantity $\gamma$
   is the relaxation rate of the order parameter within the TDGL
   theory.
A nonzero value of $\Re\gamma$ is at the basis of the
   phenomenon of paraconductivity \cite{Larkin:02}.
One finds $\Re\gamma = \pi\nu/8T$ at temperature $T$, where $\nu$ is
   the density of states.

Under complex conjugation and inversion of the magnetic field in
   Eq.~(\ref{eq:TDGL}), the equation for $\psi^\ast_\ell$ would be the same as
   that for $\psi_\ell$, provided that $\Im\gamma=0$.
Thus, a nonzero value of $\Im\gamma$ is associated with a breaking of
   electron-hole symmetry \cite{Fukuyama:71,Aronov:95}.
The condition $\Im\gamma\neq0$ then gives rise to fluctuation effects
   on the Hall conductivity \cite{Ullah:91}, the Nernst effect
   \cite{Ullah:91,Livanov:99,Ussishkin:02}, and the thermopower
   \cite{Ullah:91,Blanter:92}. 

The fluctuation contibution to several transport properties, such as
   paraconductivity, magnetoconductivity, Nernst effect,
   and thermopower, have been evaluated under several approximations
   (see Ref.~\onlinecite{Larkin:02} for a review).
From the microscopic point of view, the total fluctuation contribution
   $\Delta\sigma_{xy}$ to the Hall conductivity $\sigma_{xy}$ close to
   $T_c$ can be expressed as the sum of two terms \cite{Lang:94a}:
\begin{subequations}
\begin{eqnarray}
\label{eq:AL}
\Delta\sigma_{xy}^{\mathrm{AL}} &=& \frac{e^2}{16\hbar d}
   \frac{\sigma_{xy}^N}{\sigma_{xx}^N} \beta \frac{\pi d}{72\xi_c (0)}
   \frac{1+1/\alpha}{(1+1/2\alpha)^{3/2}} \frac{1}{\varepsilon^{3/2}}
   ,\\
\label{eq:MT}
\Delta\sigma_{xy}^{\mathrm{MT}} &=& \frac{e^2}{16\hbar d}
   \frac{\sigma_{xy}^N}{\sigma_{xx}^N} \frac{4}{\varepsilon -\delta}
   \nonumber \\
   &&\times 
   \ln \left[ \frac{\varepsilon}{\delta} \frac{1+\alpha +
   (1+2\alpha)^{1/2}}{1+\alpha\varepsilon/\delta +
   (1+2\alpha\varepsilon/\delta)^{1/2}} \right],
\end{eqnarray}
\label{eq:excessHall}
\end{subequations}
\noindent
respectively related to the Aslamazov-Larkin (AL) \cite{Aslamazov:68} and
   the Maki-Thomson (MT) \cite{Maki:89} contributions.
In Eqs.~(\ref{eq:excessHall}), $\varepsilon=\ln (T/T_c ) \approx (T-T_c
   )/T_c$ is the reduced temperature, $\alpha=2\xi_c^2 (0)/d^2
   \varepsilon$, $d$ is the interlayer spacing, $\xi_c (0)$ is the
   coherence length along the $c$ axis at $T=0$,
   $\sigma^N_{\alpha\beta}$ refer to the components of the
   conductivity tensor in the absence of fluctuations,
   $\delta=\pi\hbar/8k_{\mathrm{B}} T\tau_\phi$ is the
   MT pair breaking parameter, with $\tau_\phi$ the phase relaxation
   time of the quasiparticles, and, finally, $\beta\propto\Im\gamma$
   (Ref.~\onlinecite{Ullah:91}).
While $\xi_c (0)$ and $\delta$ can be independently determined by
   fitting analogous (AL+MT) expressions for the paraconductivity
   \cite{Ullah:91}, the parameter $\beta\propto\Im\gamma$ can be
   extracted by comparison with experimental data for the excess Hall
   effect \cite{Lang:94,Lang:94a,Lang:95,Wang:99}.
Table~\ref{tab:beta} lists values of $\beta$ for several layered
   cuprate superconductors and HTS superlattices.
One can immediately observe that $\beta$ shows a direct
   correlation with $T_c$, \emph{i.e.} $|\beta|$ increases as $T_c$
   increases, which we will discuss in more detail in
   Sec.~\ref{sec:micro}.

\begin{table}[th]
\caption{Electron-hole asymmetry parameter $\beta\propto\Im\gamma$ and 
   critical temperature $T_c$ for several layered cuprates and cuprate
   superlattices.
Tha values of $\beta$ listed here have been obtained from a fit of the 
   AL+MT corrections to conductivity and Hall conductivity,
   Eqs.~(\protect\ref{eq:excessHall}), against data for excess Hall effect.}
\label{tab:beta}
\begin{tabular}{lr@{.}lr@{.}ll}
\hline
 & \multicolumn{2}{c}{$T_c$~[K]} & \multicolumn{2}{c}{$\beta$} & \\
\hline 
YBCO/PBCO ($36$~\AA/$96$~\AA) & $68$ & $68$ & $-0$ & $0003$ &
   Ref.~\protect\onlinecite{Wang:99} \\
YBCO/PBCO ($120$~\AA/$96$~\AA) & $86$ & $33$ & $-0$ & $075$ &
   Ref.~\protect\onlinecite{Wang:99} \\
YBCO & $88$ & $55$ & $-0$ & $17$ & Ref.~\protect\onlinecite{Lang:94a} \\
Bi-2223 & $105$ &  & $-0$ & $38$ & Ref.~\protect\onlinecite{Lang:95} \\
(Bi,Pb)-2223 & $109$ & & $-1$ & &
   Ref.~\protect\onlinecite{Lang:94} \\
\hline
\end{tabular}
\end{table}

\section{Evaluation of I\lowercase{m}~$\gamma$ in the presence of an ETT}
\label{sec:micro}

From a microscopic point of view, the TDGL relaxation rate $\gamma$ in
   Eq.~(\ref{eq:TDGL}) is related to the static limit of the frequency
   derivative of the retarded polarization operator as \cite{Larkin:02}:
\begin{equation}
\gamma = i \lim_{\Omega\to0}
   \frac{\partial\Pi^{\mathrm{R}}}{\partial\Omega}.
\label{eq:master}
\end{equation}
Before the analytic continuation, the polarization operator is defined as
   ($k_{\mathrm{B}} = \hbar = 1$):
\begin{eqnarray}
\Pi(\bk,i\Omega_m;z,T) &=& \frac{T}{\mathcal{N}}
   \sum_{\bq,\epsilon_n} G^{(0)} (\bq,i\epsilon_n + i\Omega_m )
   \nonumber\\
&&\times G^{(0)} (\bk-\bq,-i\epsilon_n ),
\label{eq:polarization}
\end{eqnarray}
where $\epsilon_n$ [$\Omega_m$] are fermion [boson] Matsubara
   frequencies,
\begin{equation}
G^{(0)} (\bk,\epsilon) = \frac{1}{i\epsilon - \xi_\bk}
\end{equation}
is the Green's function for free electrons with dispersion relation
   $\xi_\bk$, and the outer sum is performed over the $\mathcal{N}$ wavevectors
   $\bq$ in the first Brillouin zone (1BZ).
Here, we specifically have in mind the 2D tight-binding dispersion relation:
\begin{equation}
\xi_\bk = -2t(\cos k_x + \cos k_y ) + 4 t^\prime \cos k_x \cos k_y
   -\mu,
\label{eq:disp}
\end{equation}
with $t$, $t^\prime$ being hopping parameters between nearest and
   next-nearest neighbors of a square lattice, respectively.
Equation~(\ref{eq:disp}) has been often employed in order to describe
   the highly anisotropic dispersion relation of the cuprates.
For $\mu=\mu_c = -4t^\prime$, the Fermi surface defined by $\xi_\bk =
   0$ has a critical form and undergoes an electronic topological transition (see
   Refs.~\onlinecite{Onufrieva:99a,Onufrieva:00,Angilella:01}).
Below, we will make use of the parameters $z=(\mu-\mu_c )/4t$, measuring the
   distance from the ETT ($z=0$), and of the hopping ratio $r=t^\prime /t$
   ($0<r<\frac{1}{2}$) \cite{Pavarini:01}.
A non-zero value of $r$ implies a breaking of electron-hole symmetry,
   with the electron sub-band width decreasing, and the hole sub-band
   width increasing of an equal amount $8rt$ (Ref.~\onlinecite{Angilella:01}).
The density of states (DOS) associated to Eq.~(\ref{eq:disp}) is
   characterized by a logarithmic singularity at $z=0$.
Such a logarithmic cusp becomes weakly asymmetric around $z=0$ in the
   case $r\neq0$ (see Appendix~\ref{app:DOS}).

With the help of standard methods \cite{AGD}, the sum over electronic
   Matsubara frequencies in Eq.~(\ref{eq:polarization}) is readily
   evaluated, and after analytic continuation to the upper complex
   plane 
\begin{equation}
\Pi^{\mathrm{R}} (\bk,\Omega) = - \lim_{\delta\to 0^+} \Pi (\bk,\Omega+i\delta),
\end{equation}
one obtains:
\begin{widetext}
\begin{eqnarray}
\Pi^{\mathrm{R}} (0,\Omega;z,T) &=& \frac{i}{2\pi}
   \frac{1}{\mathcal{N}} \sum_\bk \frac{1}{\Omega + 2 \xi_\bk +
   i\delta} \left[ 
\psi \left( \frac{1}{2} + \frac{i(\xi_\bk + \Omega)}{2\pi T} \right)
-\psi \left( \frac{1}{2} - \frac{i(\xi_\bk + \Omega)}{2\pi T} \right)
   \right. \nonumber\\
&&\qquad+ \left.
\psi \left( \frac{1}{2} + \frac{i\xi_\bk}{2\pi T} \right) - \psi
   \left( \frac{1}{2} - \frac{i\xi_\bk}{2\pi T} \right) \right],
\nonumber\\
&=& - \frac{1}{2} \frac{1}{\mathcal{N}} \sum_\bk \frac{1}{\Omega + 2 \xi_\bk +
   i\delta} \left[ \tanh \left( \frac{\xi_\bk + \Omega}{2T} \right) +
   \tanh \left( \frac{\xi_\bk}{2T} \right) \right],
\label{eq:polarizationexpl}
\end{eqnarray}
\end{widetext}
where $\psi(z)$ here denotes the digamma function \cite{AS} and $\delta$ is a
   positive infinitesimal.
Performing the frequency derivative and passing to the static limit,
   as required by Eq.~(\ref{eq:master}), one has:
\begin{equation}
\gamma = \frac{i}{8T} \frac{1}{\mathcal{N}} \sum_\bk 
\frac{1}{\xi_\bk + i\delta} F \left( \frac{\xi_\bk}{2T} \right),
\label{eq:gammaexpl}
\end{equation}
where
\begin{equation}
F(y) = \frac{1}{y} \tanh y - \frac{1}{\cosh^2 y} .
\label{eq:F}
\end{equation}

\begin{figure}[th]
\centering
\includegraphics[height=0.75\columnwidth,angle=-90]{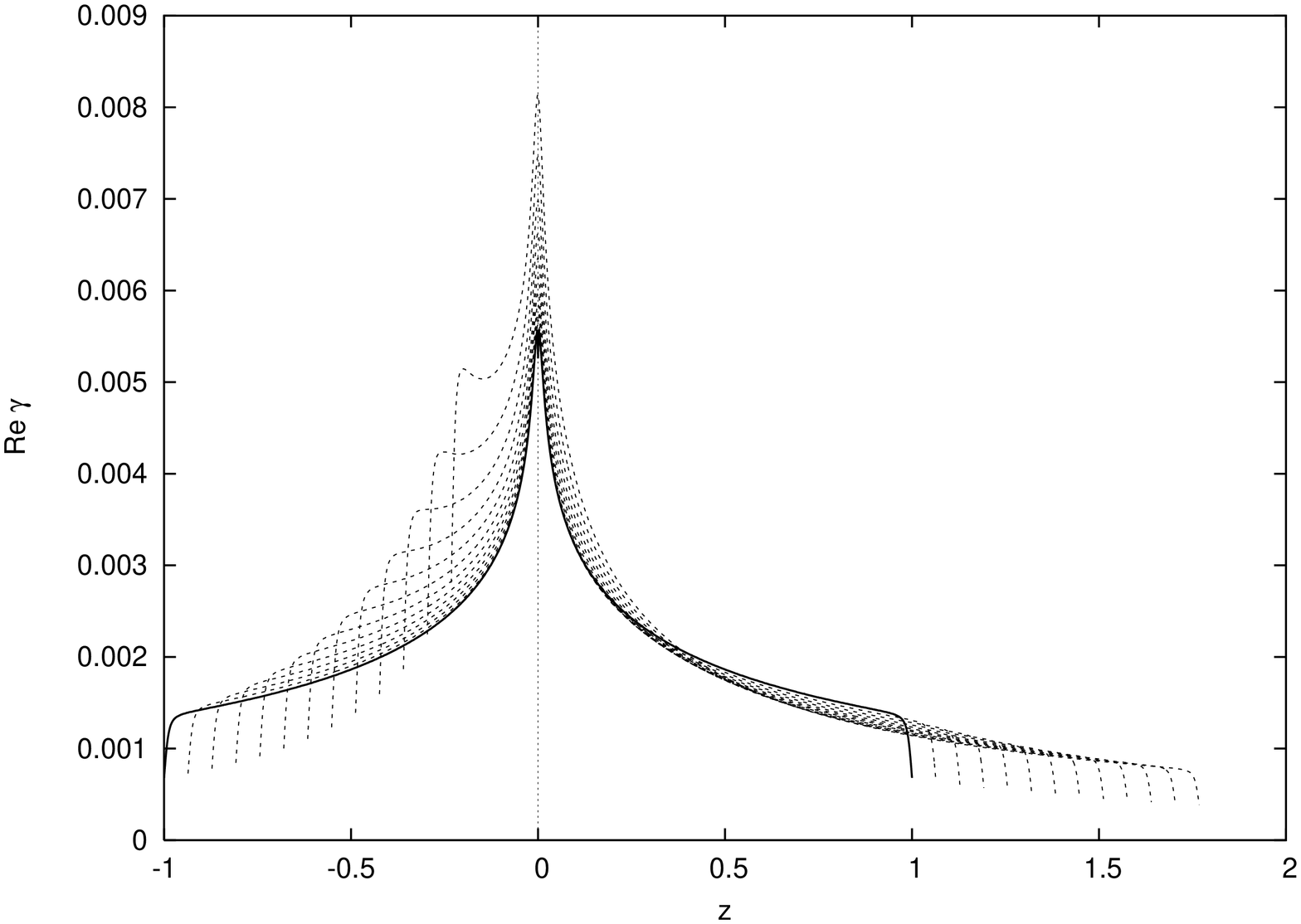}
\includegraphics[height=0.75\columnwidth,angle=-90]{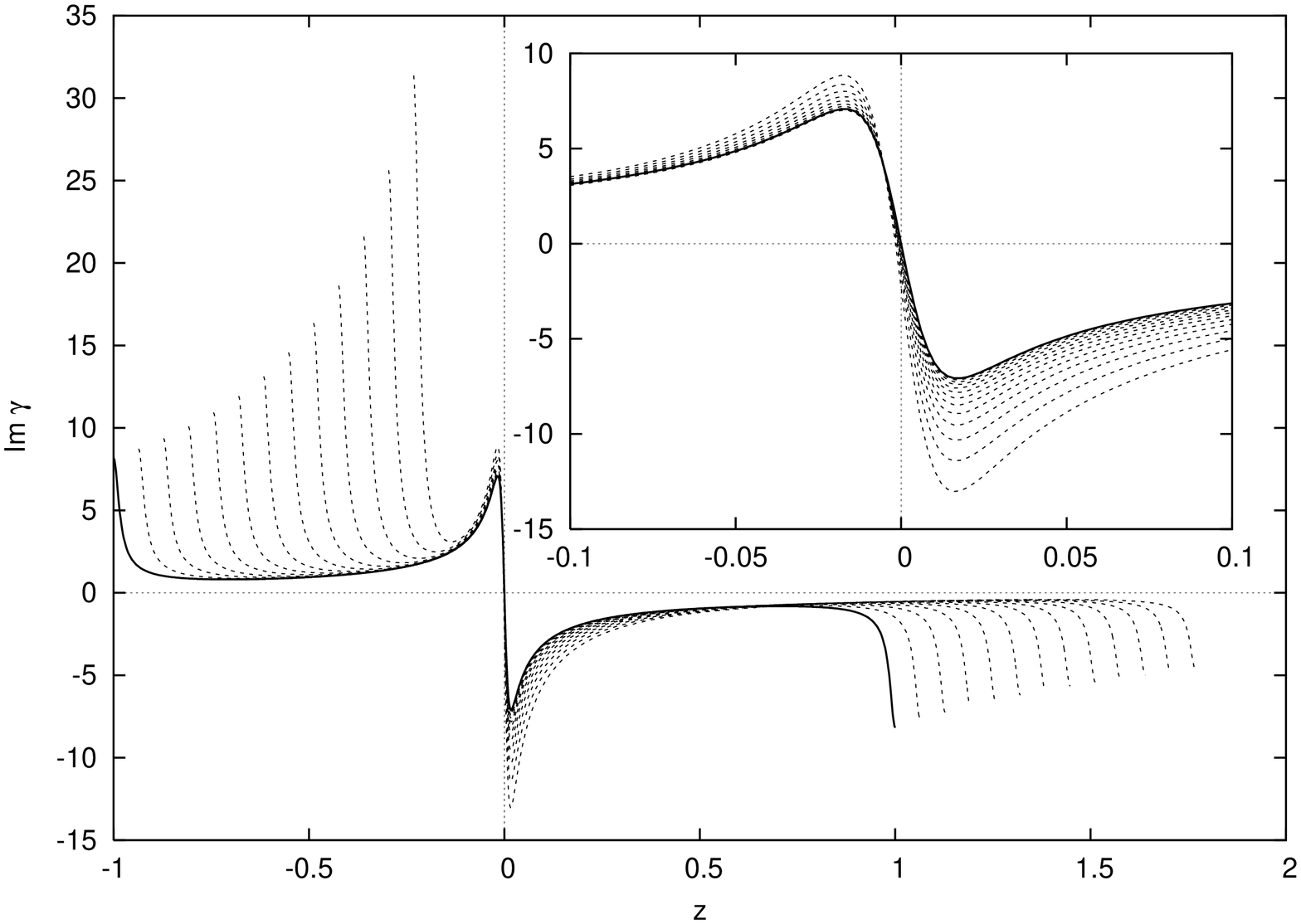}
\caption{%
Real part (top panel) and imaginary part (bottom panel) of TDGL
   relaxation rate $\gamma$, 
   Eq.~(\protect\ref{eq:gammaexpl}), as a function of the ETT parameter
   $z$ ranging over the whole bandwidth, for fixed temperature $\tau=0.005$
   and hopping ratios $r=0,\, 0.032,\, \ldots 0.384$, in units
   such that $4t=1$.
Integration in $\bk$-space in Eq.~(\protect\ref{eq:gammaexpl}) has
   been performed via the tetrahedra method, with a mesh of 125751
   $\bk$-points in the irreducible wedge of the 1BZ, and
   $\delta=10^{-6}$.
Inset in lower panel shows enlarged view of $\Im\gamma(z)$ close to the ETT.
In particular, $\Im\gamma(z)$ is an odd function of $z$ in the
   electron-hole symmetric case ($r=0$, solid line), with $\Im\gamma(z=0)=0$.
Curves corresponding to increasing values of $r$ give rise to more
   pronounced peaks in $\Im\gamma$ around the ETT point.
}
\label{fig:gammaz}
\end{figure}

Figure~\ref{fig:gammaz} shows our numerical results for the real and
   imaginary parts of the TDGL 
   relaxation rate $\gamma$ as a function of the ETT parameter $z$
   over the whole bandwidth, for a representative value of the temperature
   parameter $\tau = T/4t = 0.005$ and hopping ratios $r=0-0.384$.
As anticipated, one finds that $\Re\gamma\propto\nu(z)$, with a
   logarithmic singularity at $z=0$ and an asymmetric $z$-dependence
   in the case $r\neq0$.

\begin{figure}[th]
\centering
\includegraphics[height=0.8\columnwidth,angle=-90]{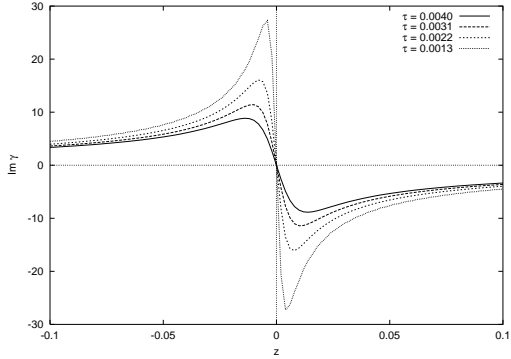}
\caption{%
$\Im\gamma(z)$ in the electron-hole symmetric case ($r=0$), for
   decreasing temperatures $\tau$ (in units such that $4t=1$).
}
\label{fig:gammat}
\end{figure}

In the electron-hole symmetric case ($r=0$), $\Im\gamma$ is an odd
   function of the ETT parameter, vanishing at $z=0$, \emph{i.e.} at
   the ETT, for all temperatures.
Close to the ETT point, $\Im\gamma$ rapidly changes sign, with two
   symmetric peaks occurring very close to the ETT point.
The height of these peaks decreases with increasing temperature
   (Fig.~\ref{fig:gammat}), and eventually diverges as $T\to0$ [see
   Eq.~(\ref{eq:diver}) below].
Such a behavior, in particular, implies a sign-changing Hall effect as
   a function of doping, and a large Hall effect close to the ETT.
Moreover, the result $\Im\gamma (z=0)=0$ is consistent with the absence
   of electron-hole asymmetry \cite{Fukuyama:71}.
A similar $z$-dependence have been demonstrated also for the
   thermoelectric power in the proximity of an ETT \cite{Blanter:92}.

\begin{figure}[th]
\centering
\includegraphics[height=0.8\columnwidth,angle=-90]{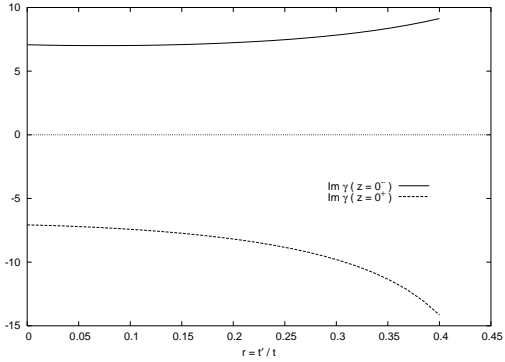}
\caption{%
Peak heights in $\Im\gamma$ around the ETT as a function of the
   hopping ratio $r=t^\prime /t$, for fixed temperature $\tau=0.005$,
   in units such that $4t=1$.
}
\label{fig:gammar}
\end{figure}

On the other hand, in the electron-hole asymmetric case ($r\neq0$),
   one in general has $\Im\gamma(z)\neq -\Im\gamma(-z)$.
However, one still recovers a sign-changing $\Im\gamma$, with
   $\Im\gamma$ vanishing very close to the ETT.
Moreover, the two peaks around the ETT have increasing heights with
   increasing hopping ratio $r$ (Fig.~\ref{fig:gammar}).
Given that a non-zero value of the hopping ratio $r$ can be associated
   with structural distortions in the $ab$ plane of the cuprates
   \cite{Angilella:02d}, one may conclude that in-plane anisotropy
   enhances the fluctuation effects associated to a non-zero value of
   $\Im\gamma$.
Moreover, on the basis of the direct correlation existing between
   $T_{c,\mathrm{max}}$ and the hopping ratio $r$
   \cite{Pavarini:01,Angilella:01}, it follows that the heights of the
   peaks in $\Im\gamma$ around the ETT increase with increasing
   $T_{c,\mathrm{max}}$ across different classes of cuprates.
Such a result is in agreement with the data listed in
   Tab.~\ref{tab:beta} for the excess Hall parameter
   $\beta\propto\Im\gamma$.

A further justification of the above numerical results can be drawn
   from an analysis of the continuum limit ($\mathcal{N}^{-1} \sum_\bk \mapsto
   \int \frac{d^2 \bk}{(2\pi)^2}$) of Eq.~(\ref{eq:gammaexpl}).
Making use of the DOS $\nu(z)$ corresponding to the dispersion relation
   Eq.~(\ref{eq:disp}) [see Eq.~(\ref{eq:appDOS}) in
   Appendix~\ref{app:DOS} below], for the imaginary part of the
   relaxation rate one obtains:
\begin{equation}
16 t^2 \Im\gamma = \frac{1}{8\tau}
\int_{-1+2r}^{1+2r} \frac{\nu(x)}{x-z} F\left(\frac{x-z}{2\tau}\right) dx,
\label{eq:Imgamma}
\end{equation}
where $\tau = T/4t$.
Eq.~(\ref{eq:Imgamma}) confirms that $\Im\gamma$ is an odd function of
   the ETT parameter $z$ in the electron-hole symmetric case, a
   source of asymmetry being provided by a non-zero value of the
   hopping ratio $r$, both through a change of the integration limits,
   and through a change in the DOS (see App.~\ref{app:DOS}).

Making use of the approximate expression fof the DOS,
   Eq.~(\ref{eq:appexp}), and of the asymptotic expansion of $F(y)$ in
   Eq.~(\ref{eq:F}),
\begin{eqnarray}
F(y) &\simeq& \frac{2}{3} y^2 , \quad |y| \leq d, \nonumber\\
           && |y|^{-1} , \quad |y| > d,
\end{eqnarray}
where $d=\sqrt[3]{3/2}$, Eq.~(\ref{eq:Imgamma}) can be integrated
   analytically, yielding the result (in units such that $4t=1$):
\begin{widetext}
\begin{eqnarray}
\Im\gamma &\simeq&
\frac{1}{2\pi^2} \frac{\ln b}{z^2 -1} \sech\left(\frac{z-1}{2\tau}\right)
   \sech\left(\frac{z+1}{2\tau}\right) \left( z \sinh \frac{1}{\tau} -
   \sinh \frac{z}{\tau} \right) +\nonumber\\
&&-\frac{1}{2\pi^2 \tau z} \left[ \ln(1-z^2) + \frac{z}{2d\tau} \ln
   \frac{2d\tau + z}{2d\tau - z} + \ln \left( 1 - \left(
   \frac{z}{2d\tau} \right)^2 \right) \right] \nonumber\\ 
&&-\frac{1}{24\pi^2 \tau^3} \left[ 2d\tau z + (4d^2 \tau^2 - z^2 ) \ln
   \frac{2d\tau + z}{2d\tau -z} \right]. 
\label{eq:Imgammax}
\end{eqnarray}
\end{widetext}
One qualitatively recovers the $z$-dependence shown in Figs.~\ref{fig:gammaz} and
   \ref{fig:gammat} for $\Im\gamma$, with $\Im\gamma$ being an odd
   function of $z$ at any given temperature $\tau$.
In particular, $\Im\gamma$ vanishes at $z=0$, where it behaves like
\begin{equation}
\Im\gamma \simeq - \frac{1}{8\pi^2 d} \frac{z}{\tau^3} ,
\end{equation}
for $|z|\ll 1$, $\tau\ll 1$.
$\Im\gamma$ is also characterized by two antisymmetric peaks occurring
   at $z\simeq \pm 2 d\tau$.
In the particle-hole symmetric case ($r=0$), the height of such peaks diverge in the
   limit $T\to0$ as
\begin{equation}
|\Im\gamma_{\mathrm{peak}} | \approx \frac{\ln 2}{2\pi^2 d}
   \frac{1}{T^2} .
\label{eq:diver}
\end{equation}
The singular behavior of $\Im\gamma$ as a function of the ETT critical
   parameter $z$ in the limit $T\to0$ is a fingerprint of quantum
   criticality \cite{Onufrieva:96,Onufrieva:99a,Onufrieva:99b,Onufrieva:00}.
In the case $r\neq 0$, additional terms arising from
   Eq.~(\ref{eq:appexp}) make this singular behavior asymmetric on the
   two sides of the ETT, as is hinted numerically by
   Fig.~\ref{fig:gammaz} (bottom panel, inset), thus showing that
   particle-hole asymmetry enhances the singular behavior of
   $\Im\gamma$ close to the ETT at low temperature.

For the sake of completeness, we also estimated $\Im\gamma$ away
   from the ETT, in the limit $|z|/\tau = |\mu-\mu_c |/T\gg 1$.
The derivation of such result is outlined in Appendix~\ref{app:away}.
Here, we just quote the final result:
\begin{equation}
\Im\gamma \approx \frac{4t\nu_0}{4T (\mu-\mu_c )} \ln \frac{2T}{|\mu-\mu_c
   |} , 
\label{eq:away}
\end{equation}
where $\nu_0 = (4t\pi^2 \sqrt{1-4r^2} )^{-1}$ is the
   density of states in the isotropic limit.
Such a result again confirms that $\Im\gamma$ is a sign-changing
   function of doping, with $\Im\gamma <0$ in the hole-like doping
   range ($z>0$).

\section{Conclusions}
\label{sec:conclusions}

We have studied the effect of superconducting fluctuations on the
   Hall conductivity of a quasi-2D layered superconductor close to an
   electronic topological transition.
Within the time dependent Ginzburg-Landau theory, such an effect is
   due to a non-zero imaginary part of the relaxation rate $\gamma$ of
   the superconducting order parameter.
We have evaluated $\Im\gamma$ as a function of the ETT parameter $z$
   and temperature, both numerically and analitically, for a quasi-2D
   dispersion relation, typical of the cuprates.
Such a dispersion is characterized by a change of topology of the
   Fermi surface at $z=0$.

In agreement with general theoretical results \cite{Fukuyama:71},
   we find that $\Im\gamma$ is a sign-changing function of the
   chemical potential, with $\Im\gamma =0$ at $z=0$ in the
   electron-hole symmetric case.
Such a result is in qualitative agreement with the universal behavior
   exhibited by the Hall anomaly in the cuprates \cite{Nagaoka:98}.

As expected, we find that $\Im\gamma$ increases
   with decreasing temperature, with a jump-like structure at the ETT whose height
   diverges as $T\to0$, and increases with increasing in-plane
   anisotropy, given by a non-zero hopping ratio $r$.
On one hand, the singular behavior developed by $\Im\gamma$ at zero
   temperature is a fingerprint of quantum criticality at $T=0$.
On the other hand, the monotonic $r$-dependence of the peak heights in
   $\Im\gamma$ at a finite temperature is in qualitative agreement with the available
   experimental results for $\Im\gamma$ in several cuprate and cuprate
   superlattices, given the direct correlation between
   $T_{c,\mathrm{max}}$ and $r$ observed for the cuprates \cite{Pavarini:01}.

\begin{acknowledgments}
We thank J. V. Alvarez, G. Balestrino, D. V. Livanov, M. Kiselev,
   P. Pfeuty, E. Piegari, P. Podio-Guidugli, and D. Zappal\`a for
   useful discussions.
\end{acknowledgments}

\appendix

\section{DOS close to the ETT in the case $r\neq0$}
\label{app:DOS}

Here, we will derive a useful asymptotic expansion of the density of
   states close to the ETT in the electron-hole asymmetric case
   ($r\neq0$).
In the following, we will employ energy units such that $4t=1$.
We start by re-writing the DOS $\nu(z)$ corresponding to the dispersion
   relation Eq.~(\ref{eq:disp}) for a square lattice (see
   Ref.~\onlinecite{Angilella:99} and refs. therein) as a function of
   the ETT parameter $z$:
\begin{equation}
\nu(z) = \frac{2}{\pi^2} \frac{1}{\sqrt{1+2r(z-r)}} K \left[
   \sqrt{\frac{1-(z-2r)^2}{1+2r(z-r)}} \right],
\label{eq:appDOS}
\end{equation}
where $K(k)$ denotes the complete elliptic integral of first kind of
   modulus $k$ (Ref.~\onlinecite{GR}).
In Eq.~(\ref{eq:appDOS}), the ETT parameter ranges as $-1+2r\leq z\leq
   1+2r$, and $\nu(z)$ is characterized by a logarithmic singularity
   at $z=0$.

In the electron-hole symmetric case ($r=0$), $\nu(z)$ is an even
   function of $z$, with $-1\leq z\leq 1$.
When electron-hole symmetry is broken by a non-zero hopping ratio $r$,
   the electron sub-band shrinks, while the hole sub-band widens of an
   equal amount $2r$, and
   the logarithmic cusp loses its symmetry around $z=0$.
In order to extract the asymptotic behavior of
   Eq.~(\ref{eq:appDOS}) around $z=0$ in the case $r\neq0$, we
   introduce the `electron' and `hole' auxiliary variables $z_1 =
   z/(1-2r)$ and $z_2 = -z/(1+2r)$, with
\begin{equation}
(1-2r)z_1 + (1+2r)z_2 = 0.
\label{eq:appcondition}
\end{equation}
Clearly, $z_1 \to z$ and $z_2 \to -z$ in the electron-hole symmetric
   case ($r=0$).
In terms of these variables, the familiar plot of the DOS
   Eq.~(\ref{eq:appDOS}) (see \emph{e.g.} Fig.~2 in
   Ref.~\onlinecite{Angilella:01}) can be seen as given by the intersection of
   Eq.~(\ref{eq:appcondition}) with the surface plot of
\begin{eqnarray}
\nu(z_1 ,z_2 ) &=& \frac{2}{\pi^2} \frac{1}{\sqrt{1-4r^2}}
   \frac{1}{\sqrt{1+z_1 + z_2}} 
\nonumber\\
&&\times
K \left[ \sqrt{ 1 + \frac{z_1
   z_2}{1+z_1 + z_2}} \right] .
\label{eq:appbiz}
\end{eqnarray}
While Eq.~(\ref{eq:appbiz}) is manifestly symmetric under particle-hole
   conjugation ($z_1 \leftrightarrow z_2$),
   Eq.~(\ref{eq:appcondition}) is particle-hole symmetric only in the
   limit $r=0$.

Expanding Eq.~(\ref{eq:appbiz}) in terms of the auxiliary variables $z_1$, $z_2$,
   and expressing the result back in terms of $z$, for $|z|\ll 1$ one
   eventually obtains 
\begin{equation}
\nu(z) \approx \frac{2}{\pi^2} \frac{1}{\sqrt{1-4r^2}} (1-az) \ln
   \frac{b}{|z|},
\label{eq:appexp}
\end{equation}
which is the desired expansion around the ETT, with $a = 2r/(1-4r^2)$
   and $b=4\sqrt{1-4r^2}$. 
A result close to Eq.~(\ref{eq:appexp}), although within the context
   of an excitonic phase, can be found in
   Ref.~\onlinecite{Kiselev:00}.
From Eq.~(\ref{eq:appexp}), one readily sees that, at lowest order in
   $z$, the source of asymmetry in the DOS logarithmic cusp at $z=0$
   comes only from the prefactor, which is linear in $z$ for $r\neq0$.

\section{Behavior of $\Im\gamma$ away from the ETT}
\label{app:away}

In the limit $|z|/\tau\gg1$, we may forget about the details of the
   dispersion relation, provided we retain its main topological
   features.
We can therefore expand Eq.~(\ref{eq:disp}) around the ETT as
\begin{equation}
\xi_\bk \approx \frac{p_1^2}{2m_1} - \frac{p_2^2}{2m_2} -z,
\end{equation}
where $p_1 = k_x$, $p_2 = k_y - \pi$, and $m_{1,2} = 2/(1\pm 2r)$ are
   the eigenvalues of the effective mass tensor around the ETT
   \cite{Angilella:01}.
Here and below, we will make use of energy units such that $4t=1$.

Our starting point will be again the general expression for the
   retarded polarization operator, Eq.~(\ref{eq:polarizationexpl}).
Passing to the new coordinates $x=p_1/\sqrt{4T m_1}$, $y=p_2/\sqrt{4T
   m_2}$, one obtains:
\begin{widetext}
\begin{equation}
\Pi^{\mathrm{R}} (0,\omega;z,T) = -\nu_0 \int_0^\infty dx \int_0^\infty
   dy \frac{\left[ \tanh(s+\omega) + \tanh(s-\omega) \right]}{s+i\delta} 
,
\label{eq:Pi1}
\end{equation}
where $s=\omega + x^2 - y^2 - 2\zeta$, $\omega = \Omega/(4T)$,
   $\zeta=z/(4T)$, and $\nu_0 = \sqrt{m_1 m_2} /(2\pi^2 ) = (\pi^2
   \sqrt{1-4r^2})^{-1}$ is the density of states in the
   isotropic case. 
According to Eq.~(\ref{eq:master}), in order to obtain $\Im\gamma$, we
   will be eventually interested in $\Re\Pi^{\mathrm{R}}$ in the limit
   $\omega\to0$.
Since the expansion in $\omega$ of the numerator in the integrand of
   Eq.~(\ref{eq:Pi1}) gives no contribution linear in $\omega$, we can
   just consider:
\begin{equation}
\Re\Pi^{\mathrm{R}} (0,\omega\to0;z,T) = -2\nu_0 \int_0^\infty dx \int_0^\infty
   dy \frac{\tanh(\omega_1 + x^2 - y^2 )}{\omega_1 + x^2 - y^2} ,
\label{eq:Pi2}
\end{equation}
where $\omega_1 =\omega-2\zeta$.
\end{widetext}

Let us now restrict to the case $z\ll -T$, in which case $\omega_1 =
   \omega + 2|\zeta| \gg1$.
The inner integral in Eq.~(\ref{eq:Pi2}) can then be estimated by
   introducing a cutoff $\delta\lesssim 1$, splitting the integration
   range in the intervals: $[0,a-\delta]$, $[a-\delta,a+\delta]$,
   $[a+\delta,\infty[$, with $a^2 = x^2 + \omega_1$, and replacing the
   hyperbolic tangent with its appropriate asymptotic expansion in
   each interval.
One eventually obtains:
\begin{equation}
\int_0^\infty dy \frac{\tanh(\omega_1 + x^2 - y^2 )}{\omega_1 + x^2 -
   y^2} \approx - \frac{1}{2\sqrt{x^2 + \omega_1}} \ln
   \frac{\delta^2}{4(x^2 + \omega_1 )}.
\label{eq:Pi3}
\end{equation}
Making use of such result back in Eq.~(\ref{eq:Pi2}), and performing
   the $\omega$ derivative as required by Eq.~(\ref{eq:master}), one
   has:
\begin{equation}
\lim_{\omega\to0} \frac{\partial
   \Re\Pi^{\mathrm{R}}}{\partial\omega}
= -\frac{\nu_0}{2} \int_0^\infty dx \frac{1}{(x^2 + 2 |\zeta|)^{3/2}}
   \ln \frac{\delta^2 e^2}{4 (x^2 + 2 |\zeta|)} ,
\end{equation}
where the last integration is trivial.
Repeating an analogous derivation in the case $z\gg T$, one eventually
   obtains the final result, Eq.~(\ref{eq:away}), to within
   logarithmic accuracy ($\delta^2 e^2 /4 \approx 1$).

\bibliographystyle{apsrev}
\bibliography{a,b,c,d,e,f,g,h,i,j,k,l,m,n,o,p,q,r,s,t,u,v,w,x,y,z,zzproceedings,Angilella}

\end{document}